\documentclass[preprint,proceedings]{rmaa}

\suppressfulladdresses 

\usepackage{paralist}

\usepackage{psfrag,color}

\def\spose#1{\hbox to 0pt{#1\hss}}
\def\gtsimm{\mathrel{\spose{\lower 3pt\hbox{$\sim$}}
        \raise 2.0pt\hbox{$>$}}}
  
\SetYear{2006}
\SetConfTitle{Massive Stars: Fundamental Parameters and Circumstellar Interactions}

\title{Particle acceleration in the colliding winds binary WR140} 

\author{
  J. M. Pittard,\altaffilmark{1} 
  and S. M. Dougherty\altaffilmark{2}}

\altaffiltext{1}{School of Physics and Astronomy, University of Leeds, Leeds, UK (jmp@ast.leeds.ac.uk).}

\altaffiltext{2}{National Research Council, Herzberg Institute for Astrophysics, Dominion Radio Astrophysical Observatory, Canada (sean.dougherty@nrc-cnrc.gc.ca).}

\shortauthor{Pittard \& Dougherty}
\shorttitle{RevMexAA(SC) Demo Document}

\listofauthors{J. M. Pittard, \& S. M. Dougherty}
\indexauthor{Pittard, J. M.}
\indexauthor{Dougherty, S. M.}

\abstract{Massive WR+O star systems produce high-temperature,
shock-heated plasma where the wind of the WR star and that of its
binary companion collide - the wind-collision region (WCR). The WCR is
a source of thermal (e.g. hard X-rays) and non-thermal
(e.g. synchrotron) emission, the latter arising from electrons and
ions accelerated to relativistic energies.  These colliding wind
binaries (CWBs) provide an excellent laboratory for the study of
particle acceleration at higher mass, photon and magnetic energy
densities than exist in SNRs.  Recent models of the non-thermal (NT)
emission from WR\thinspace140 have provided insight into this process.
}

\addkeyword{Radio Continuum: Stars}
\addkeyword{Stars: Individual: WR 140}

\begin{document}
\maketitle

\section{Introduction}
\label{sec:intro}
WR\thinspace140 (HD\thinspace193793) is the archetype of long-period
CWB systems. It consists of a WC7 star and an O4-5 star in a highly
eccentric orbit ($e \approx 0.88$), and exhibits dramatic variations
in its emission from near-IR to radio wavelengths
\citep{Williams:1990,White:1995}, and also at X-ray energies
\citep{Zhekov:2000,Pollock:2002,Pollock:2005} during its 7.9-year
orbit.  The variability appears to be linked to the WCR, which
experiences significant changes as the stellar separation varies
between 2 and 30~AU.  The orbit modulation of the synchrotron
flux has yet to be understood, but is likely due to a number of
mechanisms. The changing free-free opacity along the line of sight
through the extended stellar winds is certain to play a role, as will
the strong inverse Compton cooling of the NT electrons. In addition,
the intrinisic synchrotron luminosity and the spectral index of the NT
electron energy distribution may alter. Recently developed models of
CWBs that are based on hydrodynamical simulations of the stellar winds
and the WCR have provided a more accurate representation of the
thermal and NT emission \citep{Dougherty:2003,Pittard:2006}, and are a
first step towards distentangling these mechanims and ultimately
understanding the acceleration processes in detail.

Due to the non-unique solutions which can arise from these models, it
is essential that observations across a broad energy spectrum are
available to provide as many contraints as possible. Fortunately,
recent observations of WR\thinspace140 with the VLBA enable a full
orbit definition, including inclination, along with robust distance
and luminosity estimates \citep{Dougherty:2005}. Together, these
parameters provide essential constraints that are currently
unavailable for any other wide CWB. However, thermal X-ray and NT
X-ray and $\gamma$-ray observations are also critical to establish
some of the model parameters.

\begin{figure}[!t]
  \includegraphics[width=\columnwidth]{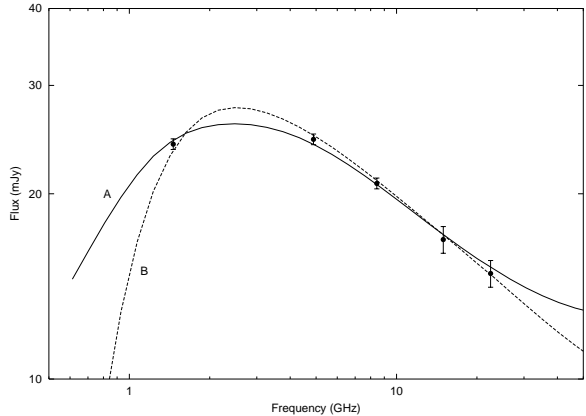}
  \caption{Model fits to radio data of WR\thinspace140 at orbital phase 0.837.
In model~A $\eta=0.22$, while model~B has $\eta=0.02$. The latter is preferred
(see text for details).}
  \label{fig:radioAB}
\end{figure}

\section{Modelling the Non-thermal Emission}
\label{sec:models}
We have applied our newly developed radiative transfer models of CWBs
to WR\thinspace140 in order to investigate the emission and
absorption processes which act to govern the radio variations. Full
details can be found in \citet{Pittard:2006b}. Rather than attempting
to model the radio lightcurve in one step, our first aim was to obtain
good fits to the radio data at phase 0.837. This phase was chosen on
the basis that the observed radio emission is close to maximum, that
an X-ray spectrum exists, and that orbital-induced curvature of the
WCR is negligible. The thermal X-ray flux is used to obtain a family
of solutions with varying WR and O-star mass-loss rates as a function
of wind momentum ratio, $\eta$. With these constraints, fits to the
radio emission (see Fig.~\ref{fig:radioAB}) lead us to favour models
with a relatively small value of $\eta$ (e.g. 0.02), on the basis that
the required line-of-sight angle is more consistent with the orbital
solution of \citet{Dougherty:2005}. Unfortunately, the VLBA images do
not have the sensitivity to directly constrain the value of $\eta$ due
to the rapid decline in the surface brightness of the WCR with
off-axis distance. In these models, free-free absorption is
responsible for the turnover between 1.6 and 5~GHz - the Razin effect
can be ruled out because it places an unacceptably large fraction of
energy into NT electrons. The post-shock B-field at the apex of the
WCR is estimated to be about 1~G, while somewhat less than 1\% of the
total available energy is transferred to NT electrons.

A key finding is that the slope of the energy spectrum of the NT
electrons, $p$, is flatter than the canonical value anticipated for
diffusive shock acceleration (i.e. $p<2$).  Several mechanisms can
produce such distributions.  The most likely are re-acceleration at
multiple shocks within the WCR \citep[e.g.][]{Schneider:1993}, and
second-order Fermi acceleration resulting from magnetic scattering off
turbulent cells \citep[e.g.][]{Scott:1975,Dolginov:1990}.  Both of
these processes could occur together if the WCR is highly structured,
as occurs naturally when clumps within the winds impact the WCR and
introduce vorticity \citep{Pittard:2007}.  Magnetic reconnection,
perhaps through resistive MHD, is another possibility.
Reconnection probably occurs throughout the volume
of the turbulent WCR, and not just at a hypothetical contact
discontinuity, and may provide additional energy for generating and
maintaining the magnetic fluctuations which drive stochastic
acceleration.

A remaining problem is that the B-field and $p$ are somewhat ill
constrained by the radio data alone. However, their degeneracy is
broken at $\gamma$-ray energies, and thus tighter constraints can be
made if future observations with GLAST are utilized. While we
generally predict lower $\gamma$-ray fluxes than previous works,
detectability should not be an issue.  However, we conclude that
WR\thinspace140 is unlikely to be detected as a TeV source with
VERITAS-4, though it may be brighter at phases closer to
periastron. Since the high stellar photon fluxes prevent the
acceleration of electrons beyond Lorentz factors $\gamma \gtsimm
10^{5}-10^{6}$, TeV emission from CWB systems will provide unambiguous
evidence of pion-decay emission from accelerated ions.

\section{Future Directions}
\label{sec:future}
While these models have provided key insights into the particle 
acceleration process(es) occuring in WR\thinspace140, much work remains.
In forthcoming work we will apply our model to other orbital phases,
and will investigate the effects of particle acceleration {\em within} 
the WCR.

\end{document}